%% file: 0_root.tex
\documentclass[sigconf, nonacm]{acmart}

\AtBeginDocument{%
  \providecommand\BibTeX{{%
    \normalfont B\kern-0.5em{\scshape i\kern-0.25em b}\kern-0.8em\TeX}}}

\acmConference[SARs: TMI - CHI Workshop]{}{April 28, 2023}{Hamburg, Germany}
\newcommand{\hidetext}[1]{} 

\acmSubmissionID{123-A56-BU3}

\begin{document}

\title{Knowing Who Knows What: Designing Socially Assistive Robots with Transactive Memory System}

\author{Yaxin Hu}
\email{yaxin.hu@wisc.edu}
\affiliation{%
  \institution{University of Wisconsin--Madison}
  \streetaddress{Dayton St}
  \city{Madison}
  \state{WI}
  \country{USA}
  \postcode{53706}
}

\author{Laura Stegner}
\email{stegner@wisc.edu}
\affiliation{%
  \institution{University of Wisconsin--Madison}
  \streetaddress{Dayton St}
  \city{Madison}
  \state{WI}
  \country{USA}
  \postcode{53706}
}

\author{Bilge Mutlu}
\email{bilge@cs.wisc.edu}
\affiliation{%
  \institution{University of Wisconsin--Madison}
  \streetaddress{Dayton St}
  \city{Madison}
  \state{WI}
  \country{USA}
  \postcode{53706}
}


\begin{abstract}
\textit{Transactive Memory System (TMS)} is a group theory that describes how communication can enable the combination of individual minds into a group. While this theory has been extensively studied in human-human groups, it has not yet been formally applied to socially assistive robot design. We demonstrate how the three-phase TMS group communication process---which involves \textit{encoding}, \textit{storage}, and \textit{retrieval}---can be leveraged to improve decision making in socially assistive robots with multiple stakeholders. By clearly defining how the robot is gaining information, storing and updating its memory, and retrieving information from its memory, we believe that socially assistive robots can make better decisions and provide more transparency behind their actions in the group context. Bringing communication theory to robot design can provide a clear framework to help robots integrate better into human-human group dynamics and thus improve their acceptance and use.
\end{abstract}


\begin{teaserfigure}
  \includegraphics[width=\textwidth]{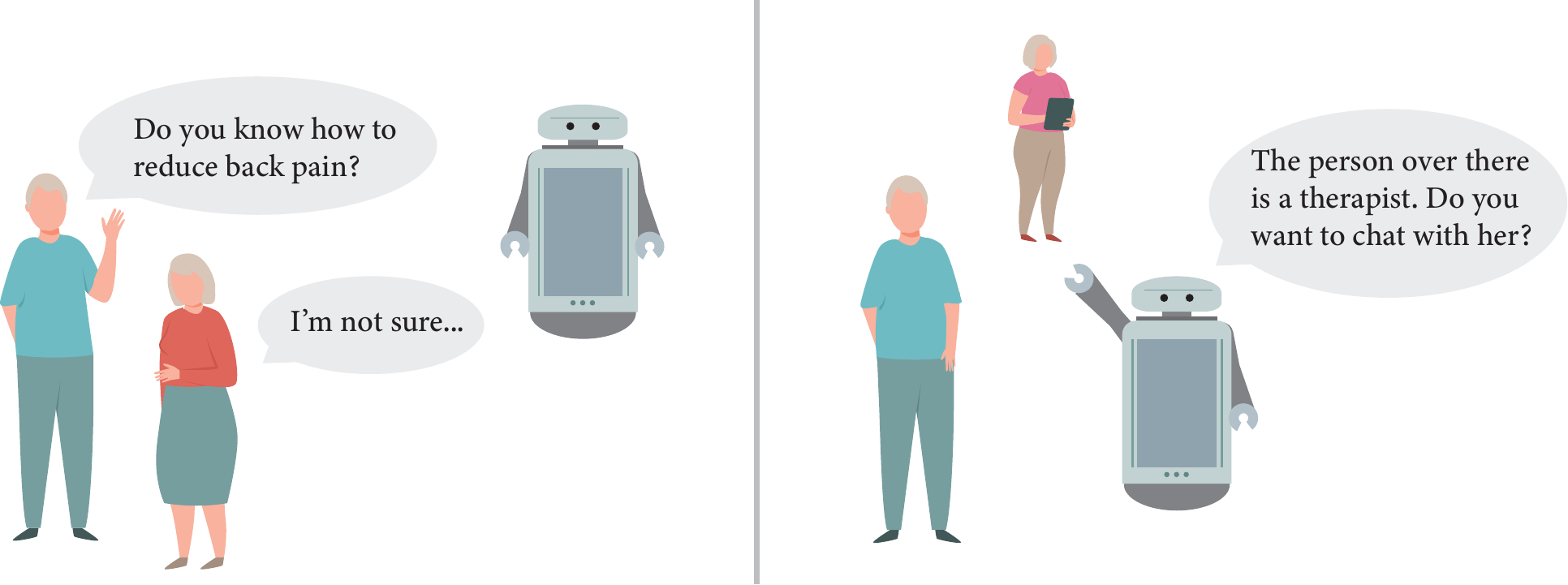}
  \caption{In this paper, we discuss how Transactive Memory System, i.e., group members knowing who knows what in the group, can be used to improve decision making in human-robot groups and lead to more transparent behaviors of socially assistive robots. We envision this framework can be used to improve group interactions by leveraging knowledge about group members' expertise. In this example, a socially assistive robot is coaching functional movement sessions for older adults. During the break, the robot observes one older adult complaining about back pain with another user (\textit{left}), it may remember meeting a therapist nearby and direct the older adult to speak to the therapist (\textit{right}).}
  \hfill \break
  \Description{Figure 1: On the left, two people are having a conversation and a robot stands behind them. The male character asks "Do you know how to reduce back pain?" and the female character answers ``I'm not sure.'' On the right, the robot is pointing to another character at the back and talking to the male character, ``The person over there is a therapist. Do you want to chat with her?''}
  \label{fig:teaser}
\end{teaserfigure}

\maketitle

\section{Introduction}
Socially assistive robots (SARs) are used to help individuals through social interactions, but do not necessarily provide any physical assistance \cite{feil2005defining, mataric2016socially, deng2019embodiment}. The goal of SARs is to provide effective and engaging interactions to assist users in areas including elderly care \cite{bemelmans2012socially, abdi2018scoping, pu2019effectiveness, robinson2014role, laura2022caregiving}, individualized learning or tutoring \cite{moro2018learning, gordon2016affective, park2019model, joe2023longterm}, physical rehabilitation \cite{fasola2012using, fasola2013socially}, cognitive disorder therapies \cite{scassellati2012robots, scassellati2007social}, \textit{etc}. The activities that SARs engage with users often include multiple stakeholders. For example, stakeholders in an educational setting could involve students, teachers, and parents, while a rehabilitation setting could involve patients, doctors, other healthcare providers, and the support system of that patient. These stakeholders naturally form group dynamics as they interact. Therefore, SARs need to fit into the group dynamics to contribute successfully to the overall goals and decision making processes that might take place in the groups.

SARs uses social cues and natural languages to interact with users and can support rapport building between the user and the robot, shift the role of the robot from being a tool to being a friend, a peer, or a companion \cite{mataric2016socially, dautenhahn2005robot}. As a result, the SAR becomes a member of the group. However, the group dynamics are more complex than simple social cues and language expressions. As the stakeholders interact as a group, they are constantly forming new connections and documenting new knowledge that impacts how they act and what decisions they make. As we introduce socially assistive robots into group dynamics, we need to consider how the robot can successfully fit into the human group context, \textit{i.e.}, considering the robot's effects on the group dynamics, group communications, and task outcomes. For example, work by \citet{tuncer2022robot} and \citet{vazquez2017towards} explored how robots can mediate the communication among human group members and affect group dynamics. However, the theories to study and explain robot behaviors in groups are understudied. 

We suggest that Transactive Memory System (TMS) is a useful group theory for understanding human-robot group dynamics. TMS has been widely studied as a group construct to explain how interpersonal communications shape the knowledge of a group. Seeing a socially assistive robot as a member of a human group, we propose to use the group theory to guide the design of robot functions and behaviors.

In this paper, we first introduce what TMS is and briefly discuss its prevalence in group communication theory. Second, we explain our motivation for applying TMS to improve decision making and trust in socially assistive robots. Then, we outline the three key phases of the group memory process---transactive memory \textit{encoding}, transactive memory \textit{storage}, and transactive memory \textit{retrieval}---and discuss the design opportunities and challenges associated with realizing each phase in a socially assistive robot. We close with a more general discussion of the potential for TMS and other formalized theories to improve decision making and transparency in SARs.



\section{What is Transactive Memory System?}


Transactive Memory System (TMS) is defined as ``a combination of individual minds and the communication among them'' \cite{wegner1985cognitive}, suggesting that people in a group depend on each other's thoughts and generate knowledge through communication processes. As a result, the overall knowledge generated by the group members is larger than adding up each individual's knowledge. 

TMS has two core components: group knowledge and communication process \cite{wegner1985cognitive, wegner1987transactive}. The first component is the knowledge of the group members, also known as ``who knows what'' \cite{lewis2007group}, suggesting that transactive memory consists of the knowledge of everyone in the group as well as the location of that knowledge among the group members. The second component of TMS refers to the communication process to form and use the transactive memory. The process is viewed as analog of individual memory with respect to three phases of knowledge processing: transactive encoding, transactive storage, and transactive retrieval. These three stages provide a framework to study how communication takes place and how it affects the knowledge processing and generation in each stage. Through the encoding, storage, and retrieval phases, the transactive memory is formed through differentiation and integration structures \cite{wegner1985cognitive, wegner1987transactive}. The differentiation structure introduces the distinctions of personal memories in the group. The integration structure is the process for group members to reach a shared understanding of the topic with initially disparate opinions. Integration could also create new perspectives or ideas that none of the group members had previously considered. 

In the same vein, \citet{hollan2000distributed} used the theory of \textit{Distributed Cognition} \cite{hutchins1995cognition, salomon1997distributed} as a theoretical framework to study human-computer interaction (HCI), suggesting that cognition is embodied and emerges from interactions among internal and external resources. While \textit{Distributed Cognition} in HCI considers a wider range of external factors including social organization, environment, and culture, TMS focuses on the social communication aspect of groups and views communication as a core component of the group memory formation. In recent years, an increasing body of work has studied the effects of technologies on TMS, such as the dynamics of team communication with the integration of an intelligent assistant \cite{yan2021communication}, cognitive processes of individuals  using the Internet \cite{firth2019online, firth2020exploring}, and the computer-mediated communications in virtual teams \cite{belbaly2016knowledge}. However, we have yet to see a discussion of TMS formally applied to socially assistive robot design.

\section{Why Use Transactive Memory System with Socially Assistive Robots?}

Group memory and knowledge is a key element in a group's decision making. Prior research has suggested that human memories are embodied, distributed, and extended in the environment as opposed to being an isolated cognitive process in the brain \cite{barnier2008conceptual, hutchins1995cognition}. Human groups naturally distribute memories to individual group members and form transactive memory through communication among the group members. 
Socially assistive robots are built with an increasing level of agency and autonomy, including better natural language capabilities that enable more natural interaction within a human group. 
These robots decide what actions to perform, and their behaviors influence the group decision making process. For example, if a robot decides to provide a reminder to the group at a specific time, it could prompt them to take more immediate action on that reminder, whereas without the reminder, they may opt to skip that task or save it for later. 

The robot can play an integral role as a group member in guiding the group based on the decisions it makes regarding which actions to perform. Therefore, if we can design robot behaviors using a clear TMS structure, we can improve the group dynamics and decision making overall. Furthermore, designing SARs with TMS can provide more explainable behaviors at each stage of the transactive memory process, thus improving the explainability of SARs. While elements of TMS are already seen organically emerging in the design of robot behaviors in a group dynamic \cite{urcola2011adapting}, we believe that introducing TMS as a formal design framework can provide structure and guidance for future researchers and designers when considering how to best design SARs for group interactions.


\section{What Would Transactive Memory System Look Like on a Robot?}
We now demonstrate how TMS can serve as a framework for SAR design, which is based on the three-step cognitive process previously introduced: encoding, storage, and retrieval. We will present each step in parallel with a working example to illustrate what it may look like when implemented on a robotic exercise coach.

\paragraph{Working Example Scenario.} The scenario used in the working example is as follows: Imagine a socially assistive robot in a nursing home that provides coaching for older adults to help them perform functional movements. Besides users in the coaching session---older adults in this case---other stakeholders involved in the activity could include staff in the nursing home, physical therapists, and other bystanders. The staff facilitate the users to join the session, the physical therapists assess the user's physical condition and assign the appropriate exercise or treatment, the family members monitor the general well being of the users, and the robot helps with delivering the coaching content, motivating the users, and assessing their performance. Considering the coaching as a group project, every group member, including the robot, plays different roles to make the activity successful. In the group's transactive memory system, the robot is the expert who has knowledge on the coaching sessions and each user's condition in the group. 

\subsection{Transactive Memory Encoding}

\paragraph{Overview. } Transactive memory encoding is the phase where the group members get to know each others' specialties and knowledge. Socially assistive robots can be designed to have conversations with other group members to understand their expertise and responsibilities. The robot should also show what it can do, have other group members understand its capabilities and build trust in this process. This interaction can serve as the foundation for building the transactive memory system of the group. While the robot is cataloging information, the other group members are doing the same, even if they are not actively aware of the process.

\paragraph{Working Example. } Imagine a new therapist joins the exercise group. How can they transition smoothly into the group and collaborate effectively? They need to get to know the expertise and responsibilities of other group members, while letting the other members know what their specialties are. This situation is one example of transactive memory encoding. The participating older adults may introduce themselves and describe why they are attending the exercise sessions; the other therapists could introduce themselves and explain their specializations and past experience with the users; the family members could explain what they hope the users will gain from the sessions. As another group member, the robot can take an active role in the process of forming transactive memory in the group. That is, the robot can share with the new therapist what its roles and capabilities are. The robot also needs to remember information about the other group members so that it can leverage that knowledge for later decision making. The robot should recognize details such as the new therapist's name, the reactions others had to them, or even whether certain group members were not present during the initial introduction.

\subsection{Transactive Memory Storage}
\paragraph{Overview. }  Transactive memory storage is the stage when the group has already formed the shared knowledge and has a directory of the expertise of each group member. The transactive memory in this phase is highly modifiable by subsequent events and communications. Socially assistive robots need to keep track of the changes of the transactive memory in the group and update its own memories. They should remember not only the key information from the memory but also details about why it was catalogued, such as when the interaction took place or who provided it with the information. In the memory storage phase, the robot needs to constantly update its knowledge base, while keeping other group members aware of the changes. For its own internal state, the robot needs to be transparent about what it is doing and what it intends to do next. The robot could increase its explainability by showing how its internal state is updated over time. 

\paragraph{Working Example. } Imagine that the robot initially saw the user in the exercise session being unfriendly to the new therapist, so it encoded some memory that it should not pressure them to work together. However, weeks later, the user asks the robot why they have not worked with the new therapist yet. The robot is able to explain that it did not encourage them to work together due to its previous observation that they did not get along. The user then tells the robot that they were looking forward to working with the new therapist, so the robot must then update its memory such that it might encourage them to work together after all. At the next session, the robot can use its updated memory.

\subsubsection{Transactive Memory Retrieval}
\paragraph{Overview. } In the transactive memory retrieval phase, the group obtains the needed information through identifying the group member with the specific expertise. The socially assistive robot may provide the information if it knows, or it may point to the other group member who has the knowledge. The robot needs to identify the member's role in the group and only share the appropriate information to that group member. 

At a lower level, the retrieval process is a coordination process among the group members. The robot needs to recall the expertise of other group members and reach out for help if needed. For example, the robot may not always be able to provide the requested help from other group members. The robot could show what it knows about each group member and what they can do. To increase the transparency in the communication, the robot could also present when they obtained this information and what the interaction looked like at the encoding time (\textit{e.g.}, what conversations occurred so the robot knows the information). The robot may also need to contact other group members to handle unexpected events. The robot would need to make decisions on whom to seek help from based on its knowledge of the group and past experiences. 

\paragraph{Working Example. } The robot observes one day that the user who is normally hard working is very weak and unmotivated in one session. The robot asks the user if they are okay, and the user says they did not sleep well and feel unwell today. Upon hearing this information, the robot decides to ask one of the therapists to decide whether or not the user should continue, since the therapist is an expert in that topic. This intervention from the robot leads to the therapist ending the user's exercise session early, possibly preventing a fall or other injury. However, at the end of the session, a bystander asks where the sick user went. The robot tells the bystander that it is not allowed to share personal information about other users.

\section{Discussion}

We presented how the TMS communication process can be translated to the design of socially assistive robots. This model can provide a cohesive framework for considering the complex dynamics of social robot interaction, especially regarding how to manage information within groups. Providing a clear structure for how the robot gains information, stores and updates information, and uses that information to make decisions can lead to better interaction design by improving group dynamics and offering better explainability behind the robot's decisions and actions. While we provide a high-level discussion of how TMS can improve socially assistive robot decision making, there remain significant design opportunities and challenges related to each step of the communication process. Below, we discuss design opportunities related to each step of the communication process.

 \paragraph{Design for Encoding. } Communication factors and strategies can contribute to the effectiveness of the encoding stage. For example, when the robot is brought to the group for the first time, group members need to learn about the robot's capabilities. One communication strategy that can be used for the robot design is ``generation effect,'' \cite{wegner1985cognitive} that people tend to remember better if they actively generate information rather than passively experience and process it. This concept suggests that the robot could provide more engaging interaction through questions and answers (Q\&A), emotional expressions, \textit{etc}., to help the group better remember what value the robot could provide and how they can interact with it. Another example of communication strategies is the use of scaffolding, \textit{i.e.}, differentiating higher-order and lower-order information \cite{wegner1985cognitive}. Higher order information is the category or sets of lower order information. For example, ``fruit'' is the higher order information while ``apple'' is lower order information \cite{wegner1985cognitive}. The robot could present the higher order knowledge with the group members during the encoding phase to lower the cognitive load of the group so that the group knows the robot's expertise without remembering the details.  

 \paragraph{Design for Storage. } Socially assistive robots can employ visualization tools to improve the transparency of the group's transactive memory and prevent errors in group decisions. For instance, it can present the change record of its knowledge through visualization, increasing the group's awareness of the current knowledge and expertise distribution. In this way, the robot helps to improve the information transparency in the group and helps the group make more informed decisions.

 \paragraph{Design for Retrieval. } Multiple design factors can contribute to the transactive memory retrieval process, such as interactive cuing, contextual information, and social climate in the group \cite{yan2021communication}. Interactive cuing has been found to benefit the retrieval process, in that one person uses a cue from others to recall the full information \cite{wegner1987transactive, hollingshead1998retrieval}. In the same vein, the socially assistive robot could help other group members recall the full information through the interactive cuing if it has partial knowledge of the event or has shared experience with the group member. Another design opportunity for transactive memory retrieval is the contextual information. The presence of the robot could provide the necessary context of the information encoding and leads to a better retrieval outcome. Furthermore, a positive social climate in the group can increase people's willingness to seek expertise from other group members and benefit the transactive memory retrieval. These benefits motivate the design of the robot to build a positive social relationship with other group members. 

\section{Conclusion}
This paper summarized the theory of Transactive Memory System (TMS) and demonstrated how TMS can be used to design socially assistive robots. We posit that explicitly incorporating TMS into robot design will further enhance their ability to engage in successful, productive human group interactions where the robot is also a group member. Moreover, TMS can be applied to design robots in multi-robot groups where each robot has their own expertise and knows the expertise of other robots as well as human stakeholders. Future study need to explore concrete implementation of TMS in the human-robot groups and evaluate the effect of TMS on the group's behaviors and decision making.







\input{8_acknowledgement}

\bibliographystyle{ACM-Reference-Format}
\bibliography{literature}

\end{document}

%% file: 8_acknowledgement.tex
\section{Acknowledgement}
The teaser image is modified from an image by macrovector on Freepik.

%% file: 0_root.bbl

\begin{thebibliography}{32}


\ifx \showCODEN    \undefined \def \showCODEN     #1{\unskip}     \fi
\ifx \showDOI      \undefined \def \showDOI       #1{#1}\fi
\ifx \showISBNx    \undefined \def \showISBNx     #1{\unskip}     \fi
\ifx \showISBNxiii \undefined \def \showISBNxiii  #1{\unskip}     \fi
\ifx \showISSN     \undefined \def \showISSN      #1{\unskip}     \fi
\ifx \showLCCN     \undefined \def \showLCCN      #1{\unskip}     \fi
\ifx \shownote     \undefined \def \shownote      #1{#1}          \fi
\ifx \showarticletitle \undefined \def \showarticletitle #1{#1}   \fi
\ifx \showURL      \undefined \def \showURL       {\relax}        \fi
\providecommand\bibfield[2]{#2}
\providecommand\bibinfo[2]{#2}
\providecommand\natexlab[1]{#1}
\providecommand\showeprint[2][]{arXiv:#2}

\bibitem[Abdi et~al\mbox{.}(2018)]%
        {abdi2018scoping}
\bibfield{author}{\bibinfo{person}{Jordan Abdi}, \bibinfo{person}{Ahmed
  Al-Hindawi}, \bibinfo{person}{Tiffany Ng}, {and} \bibinfo{person}{Marcela~P
  Vizcaychipi}.} \bibinfo{year}{2018}\natexlab{}.
\newblock \showarticletitle{Scoping review on the use of socially assistive
  robot technology in elderly care}.
\newblock \bibinfo{journal}{\emph{BMJ open}} \bibinfo{volume}{8},
  \bibinfo{number}{2} (\bibinfo{year}{2018}), \bibinfo{pages}{e018815}.
\newblock


\bibitem[Barnier et~al\mbox{.}(2008)]%
        {barnier2008conceptual}
\bibfield{author}{\bibinfo{person}{Amanda~J Barnier}, \bibinfo{person}{John
  Sutton}, \bibinfo{person}{Celia~B Harris}, {and} \bibinfo{person}{Robert~A
  Wilson}.} \bibinfo{year}{2008}\natexlab{}.
\newblock \showarticletitle{A conceptual and empirical framework for the social
  distribution of cognition: The case of memory}.
\newblock \bibinfo{journal}{\emph{Cognitive Systems Research}}
  \bibinfo{volume}{9}, \bibinfo{number}{1-2} (\bibinfo{year}{2008}),
  \bibinfo{pages}{33--51}.
\newblock


\bibitem[Belbaly and Somsing(2016)]%
        {belbaly2016knowledge}
\bibfield{author}{\bibinfo{person}{Nassim~Aissa Belbaly} {and}
  \bibinfo{person}{Autcharaporn Somsing}.} \bibinfo{year}{2016}\natexlab{}.
\newblock \showarticletitle{Knowledge sharing, transactive memory system, and
  virtual team creativity}. In \bibinfo{booktitle}{\emph{Academy of Management
  Proceedings}}, Vol.~\bibinfo{volume}{2016}. Academy of Management Briarcliff
  Manor, NY 10510, \bibinfo{pages}{15649}.
\newblock


\bibitem[Bemelmans et~al\mbox{.}(2012)]%
        {bemelmans2012socially}
\bibfield{author}{\bibinfo{person}{Roger Bemelmans}, \bibinfo{person}{Gert~Jan
  Gelderblom}, \bibinfo{person}{Pieter Jonker}, {and} \bibinfo{person}{Luc
  De~Witte}.} \bibinfo{year}{2012}\natexlab{}.
\newblock \showarticletitle{Socially assistive robots in elderly care: a
  systematic review into effects and effectiveness}.
\newblock \bibinfo{journal}{\emph{Journal of the American Medical Directors
  Association}} \bibinfo{volume}{13}, \bibinfo{number}{2}
  (\bibinfo{year}{2012}), \bibinfo{pages}{114--120}.
\newblock


\bibitem[Dautenhahn et~al\mbox{.}(2005)]%
        {dautenhahn2005robot}
\bibfield{author}{\bibinfo{person}{Kerstin Dautenhahn}, \bibinfo{person}{Sarah
  Woods}, \bibinfo{person}{Christina Kaouri}, \bibinfo{person}{Michael~L
  Walters}, \bibinfo{person}{Kheng~Lee Koay}, {and} \bibinfo{person}{Iain
  Werry}.} \bibinfo{year}{2005}\natexlab{}.
\newblock \showarticletitle{What is a robot companion-friend, assistant or
  butler?}. In \bibinfo{booktitle}{\emph{2005 IEEE/RSJ international conference
  on intelligent robots and systems}}. IEEE, \bibinfo{pages}{1192--1197}.
\newblock


\bibitem[Deng et~al\mbox{.}(2019)]%
        {deng2019embodiment}
\bibfield{author}{\bibinfo{person}{Eric Deng}, \bibinfo{person}{Bilge Mutlu},
  \bibinfo{person}{Maja~J Mataric}, {et~al\mbox{.}}}
  \bibinfo{year}{2019}\natexlab{}.
\newblock \showarticletitle{Embodiment in socially interactive robots}.
\newblock \bibinfo{journal}{\emph{Foundations and Trends{\textregistered} in
  Robotics}} \bibinfo{volume}{7}, \bibinfo{number}{4} (\bibinfo{year}{2019}),
  \bibinfo{pages}{251--356}.
\newblock


\bibitem[Fasola and Mataric(2012)]%
        {fasola2012using}
\bibfield{author}{\bibinfo{person}{Juan Fasola} {and} \bibinfo{person}{Maja~J
  Mataric}.} \bibinfo{year}{2012}\natexlab{}.
\newblock \showarticletitle{Using socially assistive human--robot interaction
  to motivate physical exercise for older adults}.
\newblock \bibinfo{journal}{\emph{Proc. IEEE}} \bibinfo{volume}{100},
  \bibinfo{number}{8} (\bibinfo{year}{2012}), \bibinfo{pages}{2512--2526}.
\newblock


\bibitem[Fasola and Matari{\'c}(2013)]%
        {fasola2013socially}
\bibfield{author}{\bibinfo{person}{Juan Fasola} {and} \bibinfo{person}{Maja~J
  Matari{\'c}}.} \bibinfo{year}{2013}\natexlab{}.
\newblock \showarticletitle{A socially assistive robot exercise coach for the
  elderly}.
\newblock \bibinfo{journal}{\emph{Journal of Human-Robot Interaction}}
  \bibinfo{volume}{2}, \bibinfo{number}{2} (\bibinfo{year}{2013}),
  \bibinfo{pages}{3--32}.
\newblock


\bibitem[Feil-Seifer and Mataric(2005)]%
        {feil2005defining}
\bibfield{author}{\bibinfo{person}{David Feil-Seifer} {and}
  \bibinfo{person}{Maja~J Mataric}.} \bibinfo{year}{2005}\natexlab{}.
\newblock \showarticletitle{Defining socially assistive robotics}. In
  \bibinfo{booktitle}{\emph{9th International Conference on Rehabilitation
  Robotics, 2005. ICORR 2005.}} IEEE, \bibinfo{pages}{465--468}.
\newblock


\bibitem[Firth et~al\mbox{.}(2019)]%
        {firth2019online}
\bibfield{author}{\bibinfo{person}{Joseph Firth}, \bibinfo{person}{John
  Torous}, \bibinfo{person}{Brendon Stubbs}, \bibinfo{person}{Josh~A Firth},
  \bibinfo{person}{Genevieve~Z Steiner}, \bibinfo{person}{Lee Smith},
  \bibinfo{person}{Mario Alvarez-Jimenez}, \bibinfo{person}{John Gleeson},
  \bibinfo{person}{Davy Vancampfort}, \bibinfo{person}{Christopher~J Armitage},
  {et~al\mbox{.}}} \bibinfo{year}{2019}\natexlab{}.
\newblock \showarticletitle{The “online brain”: how the Internet may be
  changing our cognition}.
\newblock \bibinfo{journal}{\emph{World Psychiatry}} \bibinfo{volume}{18},
  \bibinfo{number}{2} (\bibinfo{year}{2019}), \bibinfo{pages}{119--129}.
\newblock


\bibitem[Firth et~al\mbox{.}(2020)]%
        {firth2020exploring}
\bibfield{author}{\bibinfo{person}{Josh~A Firth}, \bibinfo{person}{John
  Torous}, {and} \bibinfo{person}{Joseph Firth}.}
  \bibinfo{year}{2020}\natexlab{}.
\newblock \showarticletitle{Exploring the impact of internet use on memory and
  attention processes}.
\newblock \bibinfo{journal}{\emph{International journal of environmental
  research and public health}} \bibinfo{volume}{17}, \bibinfo{number}{24}
  (\bibinfo{year}{2020}), \bibinfo{pages}{9481}.
\newblock


\bibitem[Gordon et~al\mbox{.}(2016)]%
        {gordon2016affective}
\bibfield{author}{\bibinfo{person}{Goren Gordon}, \bibinfo{person}{Samuel
  Spaulding}, \bibinfo{person}{Jacqueline~Kory Westlund},
  \bibinfo{person}{Jin~Joo Lee}, \bibinfo{person}{Luke Plummer},
  \bibinfo{person}{Marayna Martinez}, \bibinfo{person}{Madhurima Das}, {and}
  \bibinfo{person}{Cynthia Breazeal}.} \bibinfo{year}{2016}\natexlab{}.
\newblock \showarticletitle{Affective personalization of a social robot tutor
  for children’s second language skills}. In
  \bibinfo{booktitle}{\emph{Proceedings of the AAAI conference on artificial
  intelligence}}, Vol.~\bibinfo{volume}{30}.
\newblock


\bibitem[Hollan et~al\mbox{.}(2000)]%
        {hollan2000distributed}
\bibfield{author}{\bibinfo{person}{James Hollan}, \bibinfo{person}{Edwin
  Hutchins}, {and} \bibinfo{person}{David Kirsh}.}
  \bibinfo{year}{2000}\natexlab{}.
\newblock \showarticletitle{Distributed cognition: toward a new foundation for
  human-computer interaction research}.
\newblock \bibinfo{journal}{\emph{ACM Transactions on Computer-Human
  Interaction (TOCHI)}} \bibinfo{volume}{7}, \bibinfo{number}{2}
  (\bibinfo{year}{2000}), \bibinfo{pages}{174--196}.
\newblock


\bibitem[Hollingshead(1998)]%
        {hollingshead1998retrieval}
\bibfield{author}{\bibinfo{person}{Andrea~B Hollingshead}.}
  \bibinfo{year}{1998}\natexlab{}.
\newblock \showarticletitle{Retrieval processes in transactive memory systems.}
\newblock \bibinfo{journal}{\emph{Journal of personality and social
  psychology}} \bibinfo{volume}{74}, \bibinfo{number}{3}
  (\bibinfo{year}{1998}), \bibinfo{pages}{659}.
\newblock


\bibitem[Hutchins(1995)]%
        {hutchins1995cognition}
\bibfield{author}{\bibinfo{person}{Edwin Hutchins}.}
  \bibinfo{year}{1995}\natexlab{}.
\newblock \bibinfo{booktitle}{\emph{Cognition in the Wild}}.
\newblock \bibinfo{publisher}{MIT press}.
\newblock


\bibitem[Lewis(2003)]%
        {lewis2003measuring}
\bibfield{author}{\bibinfo{person}{Kyle Lewis}.}
  \bibinfo{year}{2003}\natexlab{}.
\newblock \showarticletitle{Measuring transactive memory systems in the field:
  scale development and validation.}
\newblock \bibinfo{journal}{\emph{Journal of applied psychology}}
  \bibinfo{volume}{88}, \bibinfo{number}{4} (\bibinfo{year}{2003}),
  \bibinfo{pages}{587}.
\newblock


\bibitem[Matari{\'c} and Scassellati(2016)]%
        {mataric2016socially}
\bibfield{author}{\bibinfo{person}{Maja~J Matari{\'c}} {and}
  \bibinfo{person}{Brian Scassellati}.} \bibinfo{year}{2016}\natexlab{}.
\newblock \showarticletitle{Socially assistive robotics}.
\newblock \bibinfo{journal}{\emph{Springer handbook of robotics}}
  (\bibinfo{year}{2016}), \bibinfo{pages}{1973--1994}.
\newblock


\bibitem[Michaelis et~al\mbox{.}(2023)]%
        {joe2023longterm}
\bibfield{author}{\bibinfo{person}{Joseph~E. Michaelis},
  \bibinfo{person}{Bengisu Cagiltay}, \bibinfo{person}{Rabia Ibtasar}, {and}
  \bibinfo{person}{Bilge Mutlu}.} \bibinfo{year}{2023}\natexlab{}.
\newblock \showarticletitle{"Off Script:" Design Opportunities Emerging from
  Long-Term Social Robot Interactions In-the-Wild}. In
  \bibinfo{booktitle}{\emph{Proceedings of the 2023 ACM/IEEE International
  Conference on Human-Robot Interaction}} (Stockholm, Sweden)
  \emph{(\bibinfo{series}{HRI '23})}. \bibinfo{publisher}{Association for
  Computing Machinery}, \bibinfo{address}{New York, NY, USA},
  \bibinfo{pages}{378–387}.
\newblock
\showISBNx{9781450399647}
\urldef\tempurl%
\url{https://doi.org/10.1145/3568162.3576978}
\showDOI{\tempurl}


\bibitem[Moro et~al\mbox{.}(2018)]%
        {moro2018learning}
\bibfield{author}{\bibinfo{person}{Christina Moro}, \bibinfo{person}{Goldie
  Nejat}, {and} \bibinfo{person}{Alex Mihailidis}.}
  \bibinfo{year}{2018}\natexlab{}.
\newblock \showarticletitle{Learning and personalizing socially assistive robot
  behaviors to aid with activities of daily living}.
\newblock \bibinfo{journal}{\emph{ACM Transactions on Human-Robot Interaction
  (THRI)}} \bibinfo{volume}{7}, \bibinfo{number}{2} (\bibinfo{year}{2018}),
  \bibinfo{pages}{1--25}.
\newblock


\bibitem[Park et~al\mbox{.}(2019)]%
        {park2019model}
\bibfield{author}{\bibinfo{person}{Hae~Won Park}, \bibinfo{person}{Ishaan
  Grover}, \bibinfo{person}{Samuel Spaulding}, \bibinfo{person}{Louis Gomez},
  {and} \bibinfo{person}{Cynthia Breazeal}.} \bibinfo{year}{2019}\natexlab{}.
\newblock \showarticletitle{A model-free affective reinforcement learning
  approach to personalization of an autonomous social robot companion for early
  literacy education}. In \bibinfo{booktitle}{\emph{Proceedings of the AAAI
  Conference on Artificial Intelligence}}, Vol.~\bibinfo{volume}{33}.
  \bibinfo{pages}{687--694}.
\newblock


\bibitem[Pu et~al\mbox{.}(2019)]%
        {pu2019effectiveness}
\bibfield{author}{\bibinfo{person}{Lihui Pu}, \bibinfo{person}{Wendy Moyle},
  \bibinfo{person}{Cindy Jones}, {and} \bibinfo{person}{Michael Todorovic}.}
  \bibinfo{year}{2019}\natexlab{}.
\newblock \showarticletitle{The effectiveness of social robots for older
  adults: a systematic review and meta-analysis of randomized controlled
  studies}.
\newblock \bibinfo{journal}{\emph{The gerontologist}} \bibinfo{volume}{59},
  \bibinfo{number}{1} (\bibinfo{year}{2019}), \bibinfo{pages}{e37--e51}.
\newblock


\bibitem[Robinson et~al\mbox{.}(2014)]%
        {robinson2014role}
\bibfield{author}{\bibinfo{person}{Hayley Robinson}, \bibinfo{person}{Bruce
  MacDonald}, {and} \bibinfo{person}{Elizabeth Broadbent}.}
  \bibinfo{year}{2014}\natexlab{}.
\newblock \showarticletitle{The role of healthcare robots for older people at
  home: A review}.
\newblock \bibinfo{journal}{\emph{International Journal of Social Robotics}}
  \bibinfo{volume}{6} (\bibinfo{year}{2014}), \bibinfo{pages}{575--591}.
\newblock


\bibitem[Salomon(1997)]%
        {salomon1997distributed}
\bibfield{author}{\bibinfo{person}{Gavriel Salomon}.}
  \bibinfo{year}{1997}\natexlab{}.
\newblock \bibinfo{booktitle}{\emph{Distributed cognitions: Psychological and
  educational considerations}}.
\newblock \bibinfo{publisher}{Cambridge University Press}.
\newblock


\bibitem[Scassellati(2007)]%
        {scassellati2007social}
\bibfield{author}{\bibinfo{person}{Brian Scassellati}.}
  \bibinfo{year}{2007}\natexlab{}.
\newblock \showarticletitle{How social robots will help us to diagnose, treat,
  and understand autism}. In \bibinfo{booktitle}{\emph{Robotics research:
  Results of the 12th international symposium ISRR}}. Springer,
  \bibinfo{pages}{552--563}.
\newblock


\bibitem[Scassellati et~al\mbox{.}(2012)]%
        {scassellati2012robots}
\bibfield{author}{\bibinfo{person}{Brian Scassellati}, \bibinfo{person}{Henny
  Admoni}, {and} \bibinfo{person}{Maja Matari{\'c}}.}
  \bibinfo{year}{2012}\natexlab{}.
\newblock \showarticletitle{Robots for use in autism research}.
\newblock \bibinfo{journal}{\emph{Annual review of biomedical engineering}}
  \bibinfo{volume}{14} (\bibinfo{year}{2012}), \bibinfo{pages}{275--294}.
\newblock


\bibitem[Stegner and Mutlu(2022)]%
        {laura2022caregiving}
\bibfield{author}{\bibinfo{person}{Laura Stegner} {and} \bibinfo{person}{Bilge
  Mutlu}.} \bibinfo{year}{2022}\natexlab{}.
\newblock \showarticletitle{Designing for Caregiving: Integrating Robotic
  Assistance in Senior Living Communities}. In
  \bibinfo{booktitle}{\emph{Designing Interactive Systems Conference}} (Virtual
  Event, Australia) \emph{(\bibinfo{series}{DIS '22})}.
  \bibinfo{publisher}{Association for Computing Machinery},
  \bibinfo{address}{New York, NY, USA}, \bibinfo{pages}{1934–1947}.
\newblock
\showISBNx{9781450393584}
\urldef\tempurl%
\url{https://doi.org/10.1145/3532106.3533536}
\showDOI{\tempurl}


\bibitem[Tuncer et~al\mbox{.}(2022)]%
        {tuncer2022robot}
\bibfield{author}{\bibinfo{person}{Sylvaine Tuncer}, \bibinfo{person}{Sarah
  Gillet}, {and} \bibinfo{person}{Iolanda Leite}.}
  \bibinfo{year}{2022}\natexlab{}.
\newblock \showarticletitle{Robot-mediated inclusive processes in groups of
  children: From gaze aversion to mutual smiling gaze}.
\newblock \bibinfo{journal}{\emph{Frontiers in Robotics and AI}}
  \bibinfo{volume}{9} (\bibinfo{year}{2022}), \bibinfo{pages}{8}.
\newblock


\bibitem[Urcola and Montano(2011)]%
        {urcola2011adapting}
\bibfield{author}{\bibinfo{person}{Pablo Urcola} {and} \bibinfo{person}{Luis
  Montano}.} \bibinfo{year}{2011}\natexlab{}.
\newblock \showarticletitle{Adapting robot team behavior from interaction with
  a group of people}. In \bibinfo{booktitle}{\emph{2011 IEEE/RSJ International
  Conference on Intelligent Robots and Systems}}. IEEE,
  \bibinfo{pages}{2887--2894}.
\newblock


\bibitem[V{\'a}zquez et~al\mbox{.}(2017)]%
        {vazquez2017towards}
\bibfield{author}{\bibinfo{person}{Marynel V{\'a}zquez},
  \bibinfo{person}{Elizabeth~J Carter}, \bibinfo{person}{Braden McDorman},
  \bibinfo{person}{Jodi Forlizzi}, \bibinfo{person}{Aaron Steinfeld}, {and}
  \bibinfo{person}{Scott~E Hudson}.} \bibinfo{year}{2017}\natexlab{}.
\newblock \showarticletitle{Towards robot autonomy in group conversations:
  Understanding the effects of body orientation and gaze}. In
  \bibinfo{booktitle}{\emph{Proceedings of the 2017 ACM/IEEE International
  Conference on Human-Robot Interaction}}. \bibinfo{pages}{42--52}.
\newblock


\bibitem[Wegner(1987)]%
        {wegner1987transactive}
\bibfield{author}{\bibinfo{person}{Daniel~M Wegner}.}
  \bibinfo{year}{1987}\natexlab{}.
\newblock \showarticletitle{Transactive memory: A contemporary analysis of the
  group mind}.
\newblock \bibinfo{journal}{\emph{Theories of group behavior}}
  (\bibinfo{year}{1987}), \bibinfo{pages}{185--208}.
\newblock


\bibitem[Wegner et~al\mbox{.}(1985)]%
        {wegner1985cognitive}
\bibfield{author}{\bibinfo{person}{Daniel~M Wegner}, \bibinfo{person}{Toni
  Giuliano}, {and} \bibinfo{person}{Paula~T Hertel}.}
  \bibinfo{year}{1985}\natexlab{}.
\newblock \showarticletitle{Cognitive interdependence in close relationships}.
\newblock \bibinfo{journal}{\emph{Compatible and incompatible relationships}}
  (\bibinfo{year}{1985}), \bibinfo{pages}{253--276}.
\newblock


\bibitem[Yan et~al\mbox{.}(2021)]%
        {yan2021communication}
\bibfield{author}{\bibinfo{person}{Bei Yan}, \bibinfo{person}{Andrea~B
  Hollingshead}, \bibinfo{person}{Kristen~S Alexander},
  \bibinfo{person}{Ignacio Cruz}, {and} \bibinfo{person}{Sonia~Jawaid Shaikh}.}
  \bibinfo{year}{2021}\natexlab{}.
\newblock \showarticletitle{Communication in transactive memory systems: a
  review and multidimensional network perspective}.
\newblock \bibinfo{journal}{\emph{Small Group Research}} \bibinfo{volume}{52},
  \bibinfo{number}{1} (\bibinfo{year}{2021}), \bibinfo{pages}{3--32}.
\newblock


\end{thebibliography}
